\documentclass[twocolumn, amsmath,amssymb,superscriptaddress]{revtex4}
\usepackage{dcolumn}
\usepackage{bm}
\usepackage{amsmath}
\usepackage{amsfonts}
\usepackage{graphics}
\usepackage{graphicx}
\usepackage{color}

\begin{document}

\title{Market dynamics immediately before and after financial shocks: quantifying the Omori, productivity and  Bath
laws}
\author{Alexander M. Petersen}
\affiliation{Center for Polymer Studies and Department of Physics, Boston University, Boston, Massachusetts 02215, USA}
\author{Fengzhong Wang}
\affiliation{Center for Polymer Studies and Department of Physics, Boston University, Boston, Massachusetts 02215, USA}
\author{Shlomo Havlin}
\affiliation{Minerva Center and Department of Physics, Bar-Ilan University, Ramat-Gan 52900, Israel}
\affiliation{Center for Polymer Studies and Department of Physics, Boston University, Boston, Massachusetts 02215, USA}
\author{H. Eugene Stanley}
\affiliation{Center for Polymer Studies and Department of Physics, Boston University, Boston, Massachusetts 02215, USA}
\date{\today}

\begin{abstract} 
We study the cascading dynamics
immediately before and immediately after 219 market shocks. We define the time of a market shock $T_{c}$ to be the time for which the
market volatility $V(T_{c})$ has a peak that exceeds a predetermined threshold. The cascade of high volatility  ``aftershocks" 
triggered by the ``main shock" is
quantitatively similar to earthquakes and solar flares, which have been described by three empirical laws --- the Omori
law, the productivity law, and the Bath law.  We analyze the most traded $531$ stocks in  U.S. markets during the two-year
period 2001-2002 at the $1$-minute time resolution. We find quantitative relations between  the ``main shock"
magnitude $M \equiv\log_{10} V(T_{c})$  and  
the parameters quantifying
the decay of volatility aftershocks as well as the volatility preshocks.  We also find that stocks with larger trading activity react more
strongly and more quickly to market shocks than stocks with smaller trading activity. 
Our findings characterize the typical volatility response conditional on $M$, both at the market and the individual stock scale.
We argue that there is potential utility in these three statistical quantitative relations with applications in  option pricing and volatility trading.
\end{abstract}
\maketitle

\section{Introduction}
 Financial fluctuations have been a topic of study for economists \cite{Fama,
Granger},  mathematicians \cite{Mandelbrot}, 
and physicists \cite{Econophys0,Econophys1,Econophys2,Econophys3,Econophys4,Econophys5}.
Here we study financial fluctuations using concepts developed in the field of seismology \cite{Omori1, Omori2} and 
analogies from turbulent dynamics in our description of market main
shock magnitudes in order to analyze the  dynamic response of markets to  financial shocks. 
We identify parallels between energy cascades and information cascades, and also between turbulent bursts and the
clustering of volatility \cite{MarketsTurbulence}. 
Our results demonstrate three statistical regularities which relate the volatility magnitude $M \equiv \log_{10} V(T_{c})$
 to the market response before and after market shocks.
 
Common financial  ``shocks"  are relatively smaller
in  the volatility magnitude, the duration, and the number of stocks affected, than the extremely large and infrequent financial crashes.
 Devastating financial shocks such as  Black Monday (20 October, 1987) have significant aftershocks that can last for several months, and this 
 ``dynamic relaxation" is  similar to the aftershock cascade following an earthquake \cite{OmoriLillo}. 
Here we aim to better understand market shocks over a range of $M$ values. While the previous studies  have focussed on
at most a few large crashes,
we use a large data set of 219 financial ``main shocks" observed in American markets over the 2-year period 2001-2002. 
We analyze 531 frequently traded stocks corresponding to approximately 44,000,000 volatility records at a 1-minute time
resolution. We find three quantitative relations which enable answering such questions as:
\begin{itemize} 
\item[(i)] How does the rate of
volatility aftershocks decay with time, and how do the decay parameters relate to the main shock magnitude $M$? 
\item[(ii)] How many aftershocks above a given threshold 
can be expected after a  main shock of magnitude $M$? 
\item[(iii)] What is the relation between the value of the main
shock volatility $V(T_{c})$ and the second largest aftershock (or preshock)? 
\end{itemize} These three questions have been
 studied for geophysical earthquakes, and the corresponding statistical laws are  referred to respectively as the Omori
law, the productivity law, and the Bath law. 
 
The Omori law was first investigated in the context of financial crashes by Lillo and Mantegna \cite{OmoriLillo}, who
found  a power-law relaxation of fluctuations  at a 1-min time resolution for the S\&P500 over the 100-day period 
following the Black Monday crash. Power-law relaxation of  aftershocks is also observed for long 
periods following several other medium-size crashes \cite{OmoriWeber}, and also for short periods up to several days
following U.S. Federal Reserve interest rate change announcements \cite{FOMC}. One key feature of   long-range relaxation
dynamics is the  scale-free decay of large fluctuations that is typical of a system with memory, and which is
complemented by self-similarity in the decay substructure \cite{OmoriWeber}.  

We find  similar perturbation-response dynamics in the intraday volatility (absolute return) time series for many single
stocks on numerous days, indicating that markets respond in a common way to perturbations that range in size from
everyday market fluctuations to infrequent market crashes. Interestingly, the market is very responsive  to 
Federal Open Market Committee (FOMC) news, either in the form of subtle hints from the Fed or actual rate changes
(expected or unexpected),  because Fed Target rates serve as a benchmark and barometer for both U.S. and World markets \cite{FOMC}.
 Methods have been developed to use the interplay between the U.S. Treasury Bill and the Federal Funds effective rate in order to estimate the future movement of the Federal Funds target rate \cite{Jorda1}. More complex methods to estimate  the probability of interest-rate change involves analyzing the price-movement of expiring derivative contracts \cite{Hamilton1}. 
The connection between macroeconomic factors and financial markets is a tribute to the complexity and connectivity of
economic systems. It is a further indicator that news, in addition to complex order-book dynamics, can play a
significant role in explaining the large rate of occurence of large fluctuations in markets.

Here we quantify the rate $n(|t-T_{c}|)$ of volatility shocks at time $t$ both before and after a market shock occuring at time
$T_{c}$. In order to determine  $T_{c}$, we develop a method for selecting a critical time $T_{c}$ from a set of
candidate times $\{ t_{c}\}$ for which the collective market volatility of $S$ individual stocks is above a given threshold. For 19 particular 
dates corresponding to days with FOMC announcements, we compare the
values of calculated $T_{c}$ with the reported values of $T$ analyzed in \cite{FOMC},   and we find good 
prediction of  $T$ using this method. 
After this calibration, we study the relaxation dynamics of $S = 531$ stocks, analyzing the  Omori law,
 the productivity law, and the Bath law for the dynamics both before $(t<T_{c})$ and after
$(t>T_{c})$ the main market shock. 

In Section \ref{section:Data} we discuss the data, 
the quantitative methods used to calculate $n(|t-T_{c}|)$, and define collective market movement. 
 In Section \ref{subsection:Calibration} we quantify the threshold for selecting candidate cascades and calibrate using known values
of $T_{c}$ corresponding to  FOMC meetings. In Section \ref{subsection:TcB} we describe the method for choosing $T_{c}$
from each significant cascade we identify. In Section \ref{Results} we discuss the Omori-law parameters $\alpha$ and
$\Omega$, the productivity parameter $\Pi$, and the Bath law parameter $B$. We note that both $\Pi$ and $B$ are independent of the dynamical model,
and hence do not depend on  $n(|t-T_{c}|)$, the functional form of the relaxation dynamics.  For each of the statistical laws, we compare the
results we obtain for the market average  with the results we obtain for individual stocks.

\section{Data Analyzed} 
\label{section:Data} For the two-year period 2001-2002, we analyze Trades and Quotes (TAQ) data of more than $500$
stocks listed on the NASDAQ and NYSE. In order to analyze the most important subset of stocks, we  rank each stock by
the average number of transactions per minute. We find $S=531$ stocks with an average of more than 3 transactions per
minute, $S=136$ stocks with an average of more than 10 transactions per minute, and $S=20$ stocks with an average of
more than 50 transactions per minute. Unless otherwise stated, our results  correspond to the top
$S=531$ stocks, but all results become more statistically significant for smaller subsets of more
heavily traded (bellweather) stocks. 

In this paper, we study the volatility $v_{j}(t)$ of the intraday price time series $p_{j}(t)$ for stock $j$. The intraday
volatility (absolute returns) is expressed as 
\begin{equation}
v_{j}(t) \equiv \vert \ln (p_{j}(t)/p_{j}(t-\delta t))\vert \ ,
\label{volatility}
\end{equation}
where here we choose  $\delta t = 1$ minute so that we can analyze the dynamics immediately before and immediately after market shocks. 
To compare stocks, we scale each volatility time series  by the standard deviation over the entire period
analyzed. 
We then remove the ``U"-shaped intraday trading pattern (averaged over 531 stocks) from each time series. 
This establishes a normalized volatility in units of standard deviation for all minutes during the day and for all
stocks analyzed (see Ref.~\cite{OmoriWeber}). 

We introduce a volatility threshold $q$   which defines a 
binary volatility time series $n_{j}(t)$ for each stock $j$,  which we calculate from the normalized volatility time series $v_{j}(t)$ as
\begin{equation}
\label{binary} n_{j}(t) \equiv \left\{
\begin{array}{cl}
      1  \ \ \ , \ \ \ & v_{j}(t)    \geq q\\
      0 \  \ \ , \ \ \ &  v_{j}(t)  < q \ . \\
           \end{array}\right.
\end{equation}
We find that a volatility threshold $q \equiv 3 \sigma$  is large enough to distinguish
between significant fluctuations and 
normal background activity. We also choose this value $q \equiv 3 \sigma$ to provide comparison with the analysis performed in \cite{FOMC}.
The rate $n(t)$ measures the fraction of the market exceeding $q$  at time $t$, 
\begin{equation}
n(t) \equiv \frac{1}{S} \sum_{j=1}^{S} n_{j}(t) \ .
\end{equation}
The rate $n_{j}(t)$ quantifying  the volatility of a single stock $j$ corresponds to the limit $S \rightarrow1$. We define the average market
volatility $V(t)$ similarly by
\begin{equation}
V(t) \equiv \frac{1}{S} \sum_{j=1}^{S} v_{j}(t) \ .
\end{equation}

A market shock at time $T_{c}$ may result from {\it exogenous} (external) news or {\it endogenous} herding \cite{ SornettePhysA,
SornetteEndoExo}. In many cases, the market shocks can be  linked to exogenous news using archived news feeds that
cover and summarize daily market events \cite{NYT}. In order to analyze  market dynamics  symmetrically around a market
shock at time $T_{c}$, we analyze the per unit time rate $n(|t-T_{c}|)$ 
around the time $T_{c}$.  
It has been empirically observed that the response dynamics in financial markets show a power-law decay
\cite{OmoriLillo, OmoriWeber, FOMC, ShortTermRxn, information, limitorderOmori},
\begin{equation}
n(\vert t-T_{c} \vert) \sim \alpha  \vert t-T_{c} \vert^{-\Omega} \ , 
\label{equation:rate}
\end{equation}
where $\Omega$ is called the Omori power-law exponent, $\alpha$ is the cascade amplitude,  $t<T_{c}$ corresponds to before the
main shock, and $t>T_{c}$ corresponds to after the main shock. For comparison, $n(|t-T_{c}|)$ is constant for stochastic
processes with no memory, corresponding to $\Omega \equiv  0$. Hence, for an empirical value $\Omega \approx 0$, the  rate $n(|t-T_{c}|)$
 is indistinguishable from an exponential decay  for $|t-T_{c}| < \overline t$, where $\overline t$ is the
characteristic exponential time scale. 
However, for larger values of $\Omega$, the exponential and power-law response curves are distinguishable, especially
if several orders of magnitude in $\tau$ is available.
 
Instead of analyzing $n(|t-T_{c}|)$, we perform our quantitative analysis on $N(\vert t-T_{c} \vert)$, the cumulative
number of events above threshold $q$ at time $t$ minutes,  where by definition
\begin{equation}
 N(\vert t-T_{c} \vert) =\int_{T_{c}}^{t}n(\vert t'-T_{c} \vert)dt' \sim \beta (\vert t-T_{c} \vert)^{1-\Omega} 
\label{equation:Ncum}
\end{equation}
for market co-movement and 
\begin{equation}
N^{j}(\vert t-T_{c} \vert) =\int_{T_{c}}^{t}n_{j}(\vert t'-T_{c} \vert)dt' \sim \beta_{j} (\vert t-T_{c}
\vert)^{1-\Omega_{j}} \ .
\label{equation:Ncumj}
\end{equation}
for the activity of stock $j$. We perform our regression analysis on $N^{j}(\vert t-T_{c} \vert)$ because it is less noisy
and monotonic as compared to $n_{j}(\vert t-T_{c} \vert)$.

Hence, for a given day, we calculate the cumulative time series $N^{j}(t)$ from $n_{j}(t)$ for each stock $j$, where
$t = 0$ corresponds to  the opening bell at 9:30 AM ET. For comparison, we also analyze the average market response $N(t)$
of the  $S$ stocks under consideration,
which complements the study of individual stocks. 

To demonstrate our approach, in Fig.~\ref{Ntnt} we plot  $V(t)$, $  N(t)  $ and also $N^{j}(t)$ for four single stocks on   01/11/2002, a day
when there was a large  market shock corresponding to a publicized comment by the Fed chairman Alan Greenspan concerning
economic recovery which occurred at approximately $T_{c}=255$ minutes after the opening bell.

 In order to compare the dynamics before and after the market shock, we first separate the intraday time series $N(t)$
into two time series $N_{b}(t \vert t < T_{c})$, and $N_{a}(t \vert t > T_{c})$. Then,  to treat the dynamics
symmetrically around  $T_{c}$, we define the displaced time $\tau = \vert t- T_{c}
\vert \geq 1$ as the temporal distance from $T_{c}$. As an illustration,  we plot in  Fig.
\ref{Ntau} the time series  on $01/11/2002$ as a function of  $\tau$. We then employ a linear fit  to find the $\tau$ dependence of both
$N_{b}(\tau) \equiv N(T_{c})- N(\vert t- T_{c}\vert)$ and $N_{a}(\tau) \equiv N(t-T_{c})- N(T_{c})$ on a log-log scale to estimate
the Omori power-law exponents $\Omega_{b}$ before the news and $\Omega_{a}$  after the news.
 By analogy, we define $\alpha$ to be the amplitude $\alpha = \beta (1-\Omega)$ before $T_{c}$ as $\alpha_{b}$ and
after the shock as $\alpha_{a}$.

\section{Method for Determining $T_{c}$}
\label{section:Tc}

\subsection{Calibration using FOMC announcements} 
\label{subsection:Calibration} We use $n(t)$  to quantitatively determine times $T_{c}$ in which the market is moving
together, possibly  in response to an external market shock or possibly as a result of endogenous herding. In  Fig. \ref{L1ave} we plot the average daily pattern for $\overline{n}(t)$ and the standard deviation $\sigma(t)$.
 The values of $\overline{n}(t)$ and $\sigma(t)$ are not stationary, so we remove the daily trend from $n(t)$ by defining the detrended 
quantity $n'(t) \equiv (n(t)-\overline{n}(t))/ \sigma(t)$.
In order to distinguish significant moments of market co-movement from  background fluctuations,
 we use a significance threshold which we obtain 
from the distribution of market activity over the entire data set analyzed. Hence, we analyze the quantity $x(t)$
defined as, 
\begin{equation}
x(t)\equiv n(t) \frac{n(t)-\overline{n}(t))}{ \sigma(t)} \ , 
\label{xt}
\end{equation}
which is the product of $n(t)$ and $n'(t)$. 
The value of $n(t)$ quantifies the {\it size} of the market co-movement, while $n'(t)$ quantifies the {\it significance}
of the market co-movement. 
Because  $\overline{n}(t))$ is not constant during the day, we consider the normalized quantity
$n'(t)$ in order to 
remove the intraday pattern. Then, to restrict our analysis to relatively large market co-movements, we eliminate times
toward the beginning and end of each day, 
when average market activity is lower (significant morning activity is often related to overnight news \cite{Overnight}). We 
analyze the quantity $x(t)$, which is large only if both $n(t)$ and $n'(t)$ are  large. Fig. \ref{L1b}
demonstrates   how the quantity $x(t)$ is useful for amplifying  market co-movement and provides an illustration of a
significant shock with substantial preshock and aftershock dynamics. 

We analyze the time
series $x(t)$ in order to select  the set of times $\{t\}$  of the market shocks that are large in the fraction of the market involved
(large $n(t)$) as well as significant with respect to the time in which they occur (large $n'(t)$). 
We determine a significance threshold $x_{c}$ from the probability density function (pdf) of  $x(t)$  as in 
Fig.~\ref{l1x1pdf}. 
As a null model, we shuffle the order of each intraday time series $v_{j}(t)$ and obtain a shuffled market volatility
rate $n_{sh}(t)$ for each day. This preserves the empirical pdf of $v_{j}(t)$  but removes the correlations that
exist 
in the temporal structure of $v_{j}(t)$. We also plot  $\overline{n_{sh}}(t) \approx 0.23$ in Fig. \ref{L1ave} which
corresponds to  a residual $0.23$ co-movement 
due to random fluctuations.
 We compare the pdfs for $x(t)$ and $x_{sh}(t)$ in Fig.~\ref{l1x1pdf}(b), and observe a significant divergence for
$x(t) > 1$.
 
We calibrate our method for determining $T_{c}$ from candidate cascades by using the known reported values  $T$
corresponding to Fed announcements.
 We choose the value $x_{c} = 1.0$ which reproduces with the best accuracy the values of $T$ that we provide  for
comparison in Table \ref{table:dates0102}. 
The value of $x_{c}=1.0$ results in $5,804$ minutes out of $190,000$ minutes analyzed for which $x(t)> x_{c}$, or roughly 3\%
of the  2-year period with significant market co-movement. 

\subsection{Determining $T_{c}$ from candidate cascades}
\label{subsection:TcB} In a typical trading day there are many large fluctuations, for both individual stocks and
indices such as the S\&P 500 and DOW. Analysis of the return intervals and the cross-correlations in financial time series shows that significant statistical regularities exist [25-31]. This fact is 
evident in the robust probability density function of volatility which has a stable power-law tail
 for a wide range of time scales 
ranging from 1-minute to several days \cite{Econophys0, Econophys3, companypowlaw,markettimeseries2}. We
select market cascades that are above a  ``spurious fluctuations'' threshold, which we define by randomizing the order $v_{i}(t)$.
We use the corresponding shuffled values $n_{sh}(t)$ as a proxy for background noise.

\begin{table}
\caption{  Comparison of announcement times $T$ (as reported in New York Times) with the market clustering times
$T_{c}$ calculated using a threshold $x_{c}=1.0$, cascade window $\Delta l \equiv 60$ min., and $S=136$ stocks. The value of
$x(T_{c})$ corresponds to the largest value out of all the candidate $\{x\}$ in the most significant cascade of the
particular day. Dates of 19 FOMC meetings in the 2-year period between 
 Jan. 2001 - Dec. 2002, where the Federal Funds Target rate ($R_{new}$) 
was implemented by the rate change ($\Delta R$) at ($T$) minutes after the opening bell at 9:30 AM ET [32].  The absolute
relative change $ \vert \frac{\Delta R}{R_{old}}\vert \equiv \vert \Delta R(t)/R(t-1) \vert$ has typically filled the
range between $0.0$ and $0.25$.
 Note: Date** refers to {\it unscheduled} meetings, in which the announcement time did not correspond to 2:15 PM ET
($T$ = 285 minutes).}
\begin{tabular}{@{\vrule height 10.5pt depth4pt  width0pt}cccccc}\\
\noalign{
\vskip-11pt}
\vrule depth 6pt width 0pt \textbf{\em FOMC Date}  & \textbf{ $R_{new}$ (\%)}  & $\Delta R$ & $\frac{\Delta R}{R_{old}}$
& {\em T} & \textbf{\em $T_{c}$} \\
\hline 
01/03/01** & 6 & -0.5& -0.077 & 210 & 227 \\
01/31/01 & 5.5 & -0.5& -0.083 & 285 & 290\\
03/20/01 & 5 & -0.5& -0.091 & 285 & 286\\
04/18/01** & 4.5 & -0.5&-0.100 &  90 & 88\\
05/15/01 & 4 & -0.5& -0.111 & 285 & 287\\
06/27/01 & 3.75 & -0.25& -0.063& 285 & 285\\
08/21/01 & 3.5 & -0.25& -0.067& 285 & 286\\
09/17/01** & 3 & -0.5& -0.143& 0 & 16 \\
10/02/01 & 2.5 & -0.5& -0.167& 285 & 288\\
11/06/01 & 2 & -0.5 & -0.200 & 285 & 292\\
12/11/01 & 1.75 & -0.25& -0.125& 285& 287\\
01/30/02 & 1.75 & 0 & 0.00& 285 & 289 \\
03/19/02 & 1.75 & 0 & 0.00& 285 & 293\\
05/07/02 & 1.75 & 0 & 0.00& 285 & 287  \\
06/26/02 & 1.75 & 0 & 0.00& 285 & 286 \\
08/13/02 & 1.75 & 0 & 0.00& 285& 291\\
09/24/02 & 1.75 & 0 & 0.00& 285& 291\\
11/06/02 & 1.25 & -0.5 & -0.286& 285& 286\\
12/10/02 & 1.25 & 0& 0.00& 285 & 295\\
\hline
\end{tabular}
\label{table:dates0102}
\end{table}

We find  on average approximately $12$  minutes per day above the threshold $x_{c}\equiv 1.0$. So here we 
develop a 
method for selecting the most likely  time  $T_{c}$  from all candidate times with $x(t) > x_{c}$.
 For a given day, we collect  all values of $x(t)> x_{c}$ into a subset $\{ x'(t) \}$ of size $z$. From this subset, we
divide the $z$ values into $k$ cascades $\{ x'(t) \}_{i}$, which we define as localized groups using the criterion that
a  cascade ends when the time between the last $x'$ in cascade $i$ is separated from the first $x'$ in cascade $i+1$ by
a time window greater than $\Delta l$ minutes. We next assign to each cascade group $\{ x'(t) \}_i$ a weight
equal to the sum of the $x'(t)$ values belonging to the given cascade group, and select the cascade group with the largest
weight as the most significant  cascade. Within the most significant cascade group, we choose the
time corresponding to the maximum value of $x'(t)$ as the time $T_{c}$ of the main shock. We calibrate this method using
the reported times for the 19 FOMC interest rate meeting announcements, and find that the values $\Delta l \equiv 60$
and $x_{c}=1.0$ best reproduce the known set $\{ T \}$, which we provide for comparison in
Table \ref{table:dates0102}. 

Using the parameter values $x_{c}= 1.0$ and $\Delta l \equiv 60$, we find $373$ days with market shocks, out of $495$
days
 studied. If the values of $x'(t)$ were distributed uniformly across all days, then the probability of finding $122$
days
 without one $x'(t)$ is vanishingly small, which confirms that the $x'(t)$ group together forming cascades.
  We remove all days where $T_{c}$ is within $\Delta t \equiv 90$ minutes of opening $(t=0)$ or closing $(t=390)$, and all $T_{c}$ that
occur on half-days (days before or after the 4th of July, Thanksgiving or Christmas), resulting in the data set $\{ T_{c}\}$ 
constituting  $219$ individual days. We also analyzed the subset of 156 market shocks within $\Delta t \equiv 120$ minutes  of opening  or closing 
and find analogous results as those reported here for $\Delta t \equiv 90$.

Furthermore, in order to test the dependence of the data set $\{ T_{c}^{(1)}\}$ found for the time resolution $\delta t  = 1$
minute used in this paper, 
we also compare the values of $\{ T^{(5)}_{c}\}$ and $\{ T^{(10)}_{c}\}$ found using a volatility series with $\delta t
 = 5$ min.  and $\delta t  = 10$ min. resolution, respectively (see Eq. (\ref{volatility})). For each of the 219 days with a $T_{c}$ value we calculate
the absolute difference
in the time  value $T_{c}^{(\delta t)}$ using two values of  $\delta t$. We use similar values of $x_{c}$ for each
time resolution so that the number of days with market shocks for each resolution are approximately equal. The difference in
$T_{c}^{(\delta t)}$
depends on the resolution $\delta t$ and the locality $\delta T_{c}$ associated with each market shock. The
average of the absolute differences for three values of $\delta t$ are
$ \vert  T^{(5)}_{c} - T^{(1)}_{c} \vert  = 9 $ minutes and  $ \vert  T^{(10)}_{c} - T^{(1)}_{c} \vert  = 15 $
minutes. 
We estimate the standard error for a particular time resolution $\delta T_{c}^{(\delta t)} \approx 2 \delta t$, which implies that $\delta T_{c}^{(1)} \approx 2$ min. for the 1-min. time resolution.
Hence, the use of $T_{c} \pm \delta T_{c}$  does not significantly
change the results of this paper.
In the next Section, we analyze the empirical laws that quantify  the response dynamics  both before and after
significant market shocks.

\section{Results}
\label{Results} 
The analysis performed in this paper is largely inspired by the analogies between financial market crashes and earthquakes.
 Here we identify 219
cascades that meet our volatility significance criterion and that are within $\Delta t = 90$ minutes of the open or closing of the trading day.  For this set of shocks, we analyze the dynamics over the  $90$ minute period immediately before $T_{c}$ and the  $90$ minute period immediately after $T_{c}$ using the framework developed
in earthquake research \cite{EarthquakeSolara, EarthquakeSolarb, Productivity,SmallEarthquakes,seismicrates,EarthquakeAftershocks,EarthquakeIntertimeOmoriModel}. 
Although we present  results for only $\Delta t = 90$, we also analyzed the subset of 156 shocks within $\Delta t= 120$ minutes of the beginning and ending of the trading day, and find analogous results. 
We restrict our analysis to the local time period $2 \Delta t$ within the trading day so that our results are minimally affected by overnight effects and overlapping shocks (since the frequency of shocks larger than our threshold $x_{c}$ used here is approximately one per trading day). 
Hence, our analysis of the size-dependence of market relaxation dynamics,  where we relate the cascade parameters to the  market shock magnitude $M \equiv \log_{10} V(T_{c})$, pertains to the intraday horizon of  market shocks which occur relatively frequently. 


A recent study finds significant evidence of Omori power-law relaxation both before and after
common FOMC interest rate announcements \cite{FOMC}, and uses the relationship between the overnight Effective rate and the U.S.
6-month Treasury Bill  to estimate the magnitude of the financial news shock.  The dynamics before the Fed announcements, which are regularly
scheduled and pre-announced, are consistent with market's anticipated surprise in the Fed news, while the dynamics after the announcements are
related to the market's perceived  surprise in the Fed news. The Federal Reserve uses the ``announcement effect'' \cite{announcementeffect} to  control the federal funds market. Despite the calculated monetary policy, the markets react quite strongly to the news, with large Omori-law relaxation cascades that typically correspond to the relatively large $M$ found in this paper.

Closely related to the Omori relaxation of aftershocks is the productivity law, which establishes a power-law
relationship between the number of aftershocks or preshocks that follow or precede a main shock of magnitude $M$.
  This is analogous to earthquake
analysis where the magnitude is defined as  $M \approx (2/3)\log_{10} E$ \cite{seismicrates},  where $E$ is the energy associated with the stress released by the main shock. 
We justify our analogy between market volatility $V$ and earthquake energy $E$
by comparing the cumulative distribution 
\begin{equation}
P(V>s)\sim s^{-\eta_{V}}
\label{cdfV}
\end{equation} 
of volatility  in financial markets with the cumulative distribution  
\begin{equation} 
P(E>s)\sim s^{-\eta_{E}}
\label{cdfE}
\end{equation} 
 of energy $E$ in seismic earthquakes. Both cumulative distributions are asymptotically
power laws,  with $\eta_{V} \approx 3$ \cite{companypowlaw,markettimeseries2}
and the Gutenberg-Richter law  $\eta_{E} \approx 2/3$ \cite{SmallEarthquakes}. 

 For the set of 219 market shocks we analyze, we find a wide range of $V(T_{c})$, and hence a wide range of cascade
dynamics. We analyze the dynamics only within the trading day so that we can be confident that the dynamics after and before $T_{c}$ 
are related to the market shock $V(T_{c})$. 
Also, for relatively small $V(T_{c})$, it may be difficult to distinguish cascade preshocks (aftershocks) from opening and closing effects, and volatility resulting from overnight news. Furthermore,  we only analyze the first $\Delta t \equiv 90$ minutes
of each $N_{a}(\tau)$ and $N_{b}(\tau)$ time series, so that the comparison of  productivity $P_{a,b}(\Delta t)$ is not affected by time series of variable length.

In Fig. \ref{OApdf} we plot the pdf of Omori parameter values $\Omega_{a,b}$ and $\alpha_{a,b}$ obtained from
the power-law fits  of $N_{b}(\tau)$ and $N_{a}(\tau)$. 
Figs. \ref{OApdf}(a) and \ref{OApdf}(b) show the distribution of parameter values calculated for the average market responses $  N_{b}(\tau)
$ and $  N_{a}(\tau)$ corresponding to Eq.~(\ref{equation:Ncum}), while \ref{OApdf}(c) and \ref{OApdf}(d) show the
distribution of parameter values calculated from the individual stock responses $N^{j}_{b}(\tau)$ and
$N^{j}_{a}(\tau)$. 
The pdfs for individual stock values of $\Omega$ and $\alpha$ have a larger dispersion, as the response to each market
shock is not uniform across all stocks. 
For the average market response $  N(\tau) $ in Figs. \ref{OApdf}(a) and \ref{OApdf}(b) the pdfs of $\Omega$ and $\alpha$ are shifted
to larger values for  $t>T_{c}$ as compared to $t<T_{c}$. This is indicative of the stress that can build
prior to anticipated announcements \cite{announcementeffect} and the surprise that is inherent in the news. 
Larger $\Omega$ values correspond to faster relaxation times, while larger $\alpha$ values correspond to higher activity. We also
observe $\Omega < 0$, which corresponds to particular time series in which the pre-shocks or after-shocks farther away  from the
main shock  (for large $\tau$) are dominant over the volatility cascade around $T_{c}$. The 
values of the Omori parameters we find on averageing over all market shocks are given in the figure caption.

Although there is a wide distribution of Omori parameter values when considering all 219  market shocks, there is a strong correlation
between the individual stock dynamics for a given market shock. In Fig. \ref{PortInd} we relate the values of $\alpha$
and $\Omega$ calculated for the average market response to the average and standard 
deviation of $\alpha$ and $\Omega$ calculated for individual stocks for a given $T_{c}$. The strong correlation
between these quantities over 219 different dates indicates that the dispersion in the values of $\alpha$ and $\Omega$ for
individual stocks, as demonstrated in Figs. \ref{OApdf}(a) and \ref{OApdf}(c), result from the broad range of  $V(T_{c})$ values,
and further, that the dispersion does not  result merely from the range of stocks analyzed.

In Fig. \ref{OV} we plot the relation between the magnitude $M$ of each main shock and the resulting Omori exponents
$\Omega_{a,b}$ calculated from both market $ N_{a,b}(\tau) $ and individual stock $N^{j}_{a,b}(\tau)$ response
curves. Figs. \ref{OV}(a)  and  \ref{OV}(c)  show a positive relation between $M$ and the decay exponent $\Omega_{a}$, which indicates that
the market responds faster to large shocks on the intraday time scale. Figs. \ref{OV}(b)  and  \ref{OV}(d) show a significant dispersion across
all stocks for a given date. Interestingly,  we find a crossover at $M_{\mathsf x} \approx 0.5$ above which $ \Omega_{a,b} $ 
increases sharply to positive values. 
The values $\Omega \approx 0$ for $M < M_{\mathsf x}$
correspond to a dynamical cascade $n(\tau)$ that is indistinguishable from an exponential decay. Typically, small
values of $\Omega$ correspond to stocks with relatively low trading activity
which  are less sensitive to  market shocks. 
For individual stocks, we define $M$ to be the logarithm of the largest volatility within $3$ minutes of
the
main shock $T_{c}$ measured for the average market response $ N(\tau) $. This accounts for the possibility
of a stock-specific
anticipation or delay time in the volatility as a result of the mainshock $V(T_{c})$. There is also the possibility that
a spurious value of $\Omega \approx 0$ can arise from 
a stock which has a constant (potentially large) level of activity throughout the entire time period analyzed.

In Fig. \ref{AV} we plot the relation between the magnitude $M$ and the Omori-law amplitude $\alpha_{a,b}$  for both
market $ N_{a,b}(\tau) $ and individual $N^{j}_{a,b}(\tau)$ response curves. The relation
between
 $\alpha$ and $M$ is stronger than the relation between $\Omega$ and $M$, indicating that the Omori-law amplitude has a higher
information content than the Omori-law exponent.
  The strong relation for the average market response suggests that it is possible to identify precursors of
market shocks with statistical certainty. However, since often $T_{c}$  corresponds to anticipated market news, 
the significant activity prior to the main shock is a natural biproduct of trader anticipation.
 Interestingly, we also observe a critical threshold for $M_{\mathsf x}\approx 0.5$, above which the average response amplitude
$ \alpha _{a,b}$ increases suddenly, analogous to a first order transition.

 The behavior of the market cascades  above and below $M_{\mathsf x}$ are significantly different. Below $M_{\mathsf
x}$, it is typical for  $\Omega_{a,b}$ to be negative and $\alpha_{a,b} \approx 0$, indicating that the triggering shock
$V(T_{c})$ is relatively insignificant, with relatively few  preshocks and aftershocks. In this scenario, it is possible
for a few clusters of relatively large volatility towards the end of the time series $N^{j}_{a,b}(\tau)$ to dominate in
the calculation of the best-fit parameters, producing negative values for  $\Omega_{a,b}$.
  There are also cases for both $N^{j}_{a,b}(\tau)$ and $N_{a,b}(\tau)$ for which $\Omega_{a,b} \approx 0$,
corresponding to a constant rate of preshocks and aftershocks in the time period analyzed.

However, above $M_{\mathsf x}$, the cascade around $T_{c}$ is more significant, but with some anomalous properties.
Namely, in the case of the the market response $N_{a,b}(\tau)$, there is an increasing relation between $M$ and
$\Omega_{a}$, indicating that the market responds more quickly to larger  market shocks. 
This observation is consistent with the ``semi-strong'' efficient-market hypothesis, which asserts that markets
incorporate public news approximately instantly. This observation is similar to geophysical earthquakes, where it is
observed that the value of $\Omega_{a}$ increases with $M$ for a given geographical location \cite{MagnDepOmori}. Also,
for geophysical earthquakes, it is also found that the Omori amplitude increases exponentially with $M$
\cite{SmallEarthquakes}.  In the case of individual company response $N^{j}_{a,b}(\tau)$, the values of the average
$\Omega_{a,b} \approx 0.1$  saturate above $M_{\mathsf x}$, whereas the values of $\alpha_{a,b}$ increase with $M$.
Thus, the stochastic dynamics display a non-linear relationship with $M$, consistent with a non-linear shot noise
process \cite{nonlinearshotnoise}, and are a potential avenue for future theoretical research.

 In Fig. \ref{PV} we plot the  relation between $V(T_{c})$ and the productivity $P_{a}(\Delta t)$
(or $P_{b}(\Delta t)$), defined as the cumulative number of aftershocks (or preshocks) greater than the threshold $q\equiv3$ within 
$\Delta t\equiv90$ minutes of $T_{c}$.  
Motivated by the power-law relationship observed for earthquakes we fit the relations $P_{a}(\Delta t) \sim
M^{\Pi_{a}}$  and $P_{b}(\Delta t) \sim
M^{\Pi_{b}}$, and find statistically significant values for the market response $\Pi_{b} = 0.38 \pm 0.07$ and $\Pi_{a} = 0.48 \pm
0.04$, and for individual stocks $\Pi_{b} = 0.23 \pm 0.01$ and  $\Pi_{a} = 0.25 \pm 0.01$.
For earthquakes,  \cite{SmallEarthquakes} reports a range of $\Pi_{a}\approx 0.7-0.9$ values that are larger than observed here for financial markets, meaning that the productivity of physical
earthquakes increases ``faster" with main shock magnitude than does the productivity of market shocks.  Since for earthquakes $\Pi_{a}< b$, where $b\approx 1$ is the scaling exponent in the Gutenberg-Richter law, 
 this inequality establishes the relative importance of
small fluctuations as compared to large fluctuations  \cite{SmallEarthquakes}. In other words, this inequality  indicates that
small earthquakes play a larger role than large earthquakes in producing the observed number of large earthquake shocks. 
Using an analogous argument for market volatility, since the cumulative distribution exponent $\eta_{V} \approx 3$ is found to be robust across many markets
\cite{companypowlaw,markettimeseries2}, then the total number $N_{Tot}(V)$ of aftershocks
triggered by all main shock of size $V$ scales as, 
\begin{equation}
N_{Tot}(V) = P(V)P_{a}(\Delta t) \sim 10 ^{(\Pi_{a} - \eta_{V}) \log V} \ ,
\end{equation}
is a decreasing function of $V$. Hence, we  also find that aftershock cascade triggering is controlled by the contributions of the more numerous small shocks. Thus, the medium-sized market shocks (analyzed here) play a larger role than the large market shocks
 in producing the observed heavy-tailed distribution of market shocks. We further note that the
productivity is a combination of the relationships of both $\alpha$ and $\Omega$ with $V(T_{c})$, which  can be written as 
\begin{equation}
P_{a}(\Delta t) \equiv N_{a}(\Delta t) \sim (\Delta t)^{1-\Omega_{a}} \  \alpha_{a}/(1-\Omega_{a}) \sim
V(T_{c})^{\Pi_{a}} \ ,
\end{equation} 
with equivalent relation before the shock for $P_{b}(\Delta t)$.

In Fig. \ref{AOPba} we plot the values $\Omega$, $\alpha$, and $P(\Delta t)$, both before and
after the main shock at time $T_{c}$. Surprisingly, while there is little statistical relation between $\Omega_{b}$ and
$\Omega_{a}$, there is a strong relation between $\alpha_{b}$ and $\alpha_{a}$ as well as  between $P_{b}(\Delta t)$ and 
$P_{a}(\Delta t)$, for both $\Delta t = 90 $ and $\Delta t = 120$ minutes. This result could be of interest
for volatility traders and options traders who would like to anticipate the 
market dynamics {\it after} an announcement, given the dynamics {\it before} the announcement.

In Fig. \ref{Bath} we  relate the size of the largest shock $V_{1}\equiv V(T_{c})$ to the sizes
of the second largest shock $V_{2}$, both before and after $T_{c}$. The Bath law parameter $B$ quantifies the relation 
between $V_{1}$ and $V_{2}$ as
\begin{equation}
M_{1}-M_{2}=\log V_{1} -\log V_{2} = B \ .
\end{equation}
This functional form implies the relation
\begin{equation}  
V_{2}/V_{1} = C_{B} \ 
\end{equation}
and hence $B= -\log_{10} C_{B}$. Fig.~\ref{Bath}(c) is a scatter plot of  $V_{1}$ and $V_{2,a}$ which shows a linear relation  corresponding to $B_{a}=  -\log_{10}(0.90) = 0.046 $. 
Surprisingly, Fig.~\ref{Bath}(a)  also shows a strong  relation between $V_{1}$ and $V_{2,b}$ with $B_{b}= -\log_{10}(0.81) =  0.092$.
Comparing the values of $B_{b}$ and $B_{a}$, the difference between the $V_{1}$ and $V_{2}$ is smaller after $T_{c}$ than before $T_{c}$. Interestingly, both $B_{b}$ and $B_{a}$ are significantly smaller than the value $B_{E} \approx 1.2$ observed for earthquake aftershocks \cite{Productivity}, 
meaning that the largest preshock and aftershock are of comparable magnitude to the main shock. 
This significant difference between earthquakes and market shocks is largely due to the relative probabilities of observing first and second-largest events $x_{1}$ and $x_{2}$.
The conditional probability $P(x_{1}\vert x_{2}) = P(x_{1}>x_{2})$  is given by the corresponding cumulative distribution function.
 Hence, using Eq.~(\ref{cdfV}) and Eq.~(\ref{cdfE}), the ratio of the conditional probabilities for $E_{1}$ and $V_{1}$ is
\begin{equation}
\frac{P(V_{1}\vert{V_{2}})}{P(E_{1}\vert E_{2})}  = \frac{P(V_{1}>{V_{2}})}{P(E_{1}> E_{2})} \sim \frac {V_{2}^{-3}}{E_{2}^{-2/3}}  \ ,
\end{equation}
which roughly explains the $10^{2}$ factor difference $B_{E} \approx 10^{2} B_{V}$.

We also compare the volatilities  $V_{1}$ and $V_{2}$ for individual stocks in Fig.~\ref{Bath}(b) before $T_{c}$  and in Fig.~\ref{Bath}(d) after
$T_{c}$. We compute the average value $\langle V_{2} \rangle$ for linear bins, and find  $V_{1} >  \langle V_{2} \rangle $ for $V_{1} > 20$, both before and after $T_{c}$. 
Also,  Fig.~\ref{Bath} shows that  $ \langle V_{2,a} \rangle  >
\langle V_{2,b} \rangle$ for most values of $V(T_{c}$). Hence, the reaction to surpise causes larger volatility fluctuations than the anticipation of surprise.

We  further ask the question, how do the response parameters analyzed here depend on the variations between individual
stock trading patterns? To answer this question, we quantify the trading capacity of each stock by $\langle \omega \rangle$, the average number
of transactions per minute, with $3 \leq \langle \omega \rangle \leq 163$ for the $S=531$ stocks analyzed. We hypothesize that $\langle \omega \rangle$ is closely related to firm size and market impact.
 Fig.\ref{POAV}(a) shows that $\langle \alpha \rangle, \langle \omega \rangle$ and $\langle P(\Delta t) \rangle$   increase with $\langle \omega \rangle$ after $T_{c}$, indicating that stocks with a
large trading base respond to market shocks with large volatility $\langle v(T_{c}) \rangle$ (shown in Fig.\ref{POAV}(b)), but also relax more quickly, corresponding to larger $\Omega$
values. 
However, we find no statistical relation  between $\langle \omega \rangle$ and  $ \langle v_{2,a} \rangle$.
Interestingly, Fig. \ref{POAV}(b) shows that this positive relation also  applies to the dynamic response
parameters before $T_{c}$. 

\section{Discussion
\label{section:Disc}} 
Cascading avalanche dynamics are a common phenomena in complex systems ranging in scale from solar flares
\cite{EarthquakeSolara, EarthquakeSolarb} and earthquakes \cite{Omori1,Omori2, seismicrates, EarthquakeAftershocks} to microscopic vortices
in turbulent fluids \cite{turbulence}. Similar bursting phenomena is also observed in human organs, such as the heart
\cite{heart1, heart2}, lungs \cite{lung1, lung2}, and brain \cite{NeuralAvalanche1, NeuralAvalanche2, NeuralAvalanche3}, and also for common
social \cite{SornettePNAS,email,bursting, humanactivity} and economic systems \cite{MarketsTurbulence, OmoriLillo, OmoriWeber,
information, ShortTermRxn, limitorderOmori, FOMC, Earnings, VolPriceRxn}. Neural avalanches in the brain are frequent
even in the resting state, and are a signature of healthy brain functioning within the neural network. In fact, the
ability to process and disseminate information is largely attributed to the network structure of neuronal correlations
which, if inhibited by disease, lead to altered disfunctional states such as in the case of schizophrenia. Extending by
analogy, the frequency of cascades in financial markets could also be viewed as a ``healthy'' optimal state for
processing information and eliminating arbitrage among the many the degrees of freedom.
Recent work  \cite{ps10}  on the switching dynamics around  highs and lows in  finanancial time series shows further evidence of
Omori power-law scaling before and after microtrend extrema, in analogy to the market shocks at $T_{c}$ developed here. 
Interestingly, this work on switching dynamics finds cascading trends on  time scales ranging from seconds to hundreds of days.

Financial markets are subject to constant information flow, resulting in a large rate of significant events, such as 
 Fed announcements \cite{FOMC, announcementeffect}, 
quarterly earnings, splits and dividends announcements, mergers and acquisitions, institutional reports. This
information can arrive as ``expected" or come as a ``surprise". Interestingly, there are precursors 
extending more than a day in advance of expected announcements such as earnings announcements \cite{Earnings}.
Economists have long been interested in the interplay between informed and uninformed traders, and the dissemination of
information across a market consisting of rational agents. 
Early work focusses on the relationship between trading volume and price change, and the relationship between these
quantities and the qualitative notions of surprise, importance, and precision of the information \cite{VolPriceRxn}. 

Using methods from statistical physics [4,5,34] and geophysics, we analyze the absolute returns of price because of the
long-memory property, and the universality of fluctuations in this quantity  across diverse markets \cite{markettimeseries1a,markettimeseries1b,
markettimeseries2}.   Ref.\cite{VolPriceRxn} postulates that price changes reflect the average change in market
expectations, whereas trading volume reflects idiosyncratic reactions across all traders.  Recent work further quantifies
trading volume fluctuations  and finds that they are similar to price fluctuations, and furthermore, finds
significant cross-correlation between volume change and price change \cite{BorisCC}. Omori relaxation dynamics are also shown for trading volume  in
\cite{FOMC}. Here we also observe significant volume
cascading as evident in Fig. \ref{L1b}. The analysis of volume and transaction dynamics is an avenue of future research,
and could highlight the relationship between volume and price fluctuations by studying their correlation around market shocks.

To summarize, we analyzed the cascade dynamics of price volatility, which has potential applications in options pricing 
 and the pricing of other derivatives. The Black-Scholes equation in its simple form assumes that the fluctuations in the price
are constant during the duration of the option \cite{QuantumFinance}. However, more sophisticated methods
\cite{BlackScholes} incorporate time dependent price volatility, and are more realistic descriptions of the
non-stationarity of financial time series. The results in this paper are of potential interest
 for traders modeling derivatives on short time scales around expected market shocks, e.g earnings reports. The
statistical regularity of both market and individual stock behavior before and after a market shock of magnitude
$M\equiv \log_{10} V(T_{c})$ provides information that could be used in hedging, since we observe a crossover in the cascade
dynamics for $M \approx 0.5$. Knowledge of the Omori response dynamics provides a time window over which
aftershocks can be expected. Similarly, the productivity law provides a more quantitative
value for the number of aftershocks to expect. Finally, the Bath law provides conditional expectation of the largest
aftershock and even the largest preshock, given the size of the main shock. Of particular importance, from the
inequality of the productivity law scaling exponents and the  pdf scaling exponent for price volatility, we find
that the role of small fluctuations is larger than the role of extremely large fluctuations in accounting for the
prevalence of aftershocks.

\acknowledgements
We thank 
L. de Arcangelis for the idea of  investigating the Bath and productivity laws, and 
K. Yamasaki and A. Ralph for helpful suggestions, and NSF, DTRA, and ONR for financial support.

\newpage
\begin{widetext} 

\begin{figure}
\centering{\includegraphics[width=0.65\textwidth]{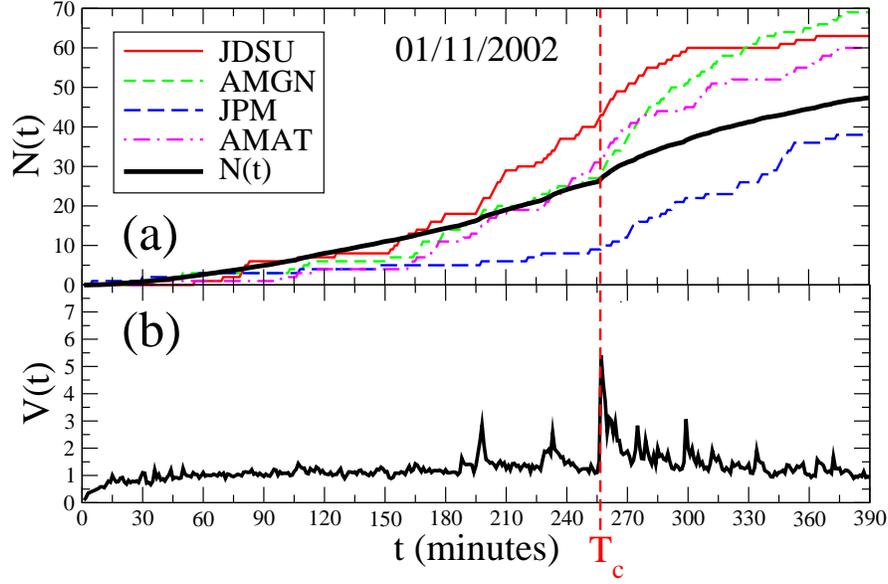}}
\caption{\label{Ntnt} (color online) Typical volatility curves on 01/11/2002 with market shock at $T_{c} = 256$ minutes. (a) The
cumulative volatility $N^{j}(t)$ for the stock of several large companies have varying behavior before $T_{c}$, but each stock shown
begins to cascade soon after $T_{c}$. The market average $ N(t) $ over all $S=531$ stocks analyzed demonstrates a
distinct change in curvature at $t=T_{c}$. (b) The average market volatility $V(t)$ demonstrates a sharp peak at
$T_{c}$, and also two precursor events at $t\approx 190$ and $\approx 230$ minutes. } 
\end{figure}

\begin{figure}
\centering{\includegraphics[width=0.65\textwidth]{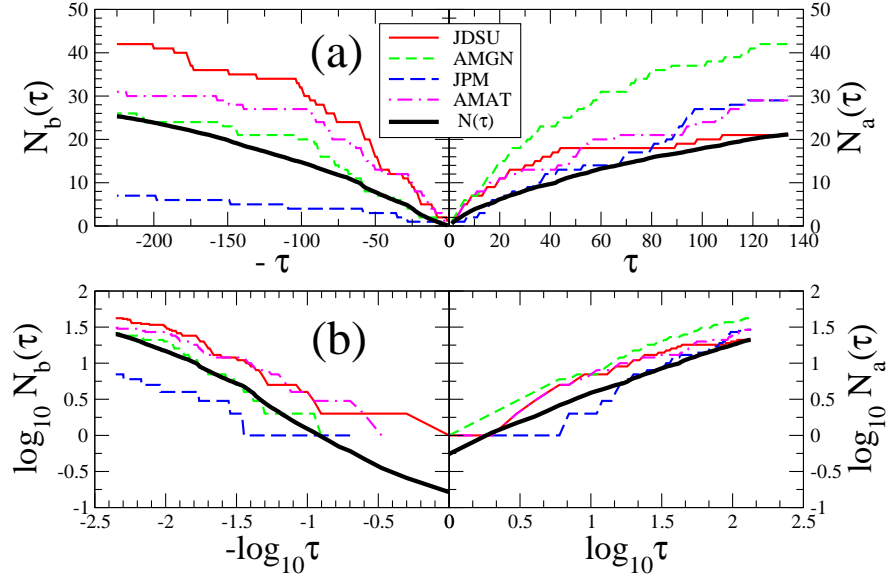}}
\caption{\label{Ntau}(color online)  (a) An illustration of $N_{b}(\tau)$ and $N_{a}(\tau)$ for the same set of curves plotted in
Fig.~\ref{Ntnt}. The 
displaced time $\tau=|t-T_{c}|$ is defined symmetrically around $T_{c}=256$ minutes on 01/11/2002.  (b) $\log_{10}
N_{b}(\tau)$ and $\log_{10} N_{a}(\tau)$ are  linear with $\log_{10} \tau$ over two orders of magnitude on a logarithmic scale. The
Omori parameters in Eq.~(\ref{equation:rate}) calculated from $N(t)$ 
are $\Omega_{b}= 0.09 \pm 0.01$, $\alpha_{b} = 0.21 \pm 0.01$ and $\Omega_{a}= 0.32 \pm 0.01$, $\alpha_{a} =
0.81 \pm 0.01$ .}
\end{figure}

\begin{figure}
\centering{\includegraphics[width=0.65\textwidth]{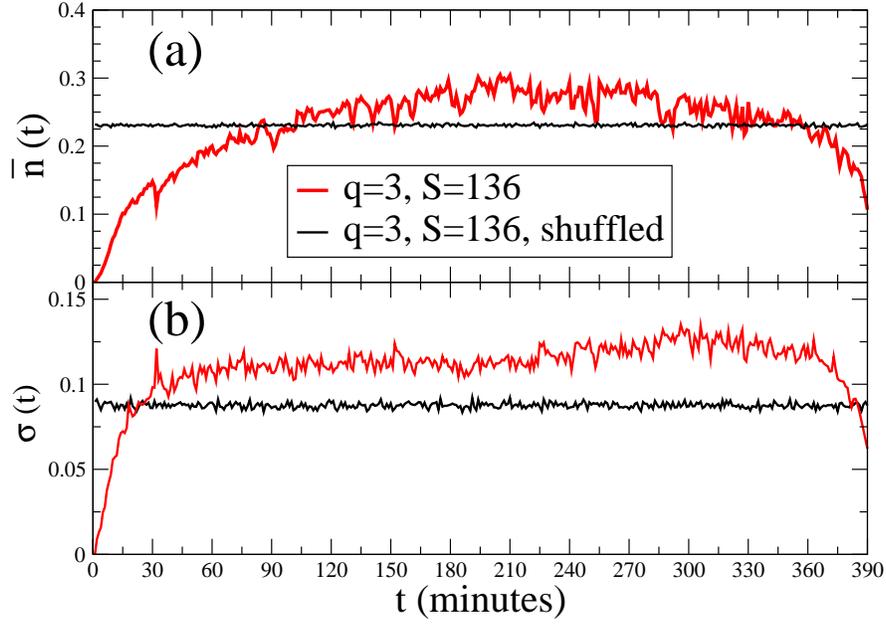}}
\caption{\label{L1ave} (color online)   The fraction $n(t)$ of the market  above the volatility threshold $q $ is non-stationary
through the trading day. We plot in (a) the average daily trading pattern $ \overline{n}(t) $  for $S=136$ stocks
and in (b)
the corresponding standard deviation, to demonstrate the trends we remove in the normalized quantity $n'(t)$. In
practice, we use the smoothed average of these curves in order to diminish statistical fluctuations on the
minute-to-minute scale. For comparison, we compute $\overline{n}_{sh}(t)  \approx 0.23$ and $\sigma_{sh} \approx 
0.09$ for shuffled $v_{i}(t)$. The values of $\overline{n}(t)$ provide an estimate for the background market
co-movement that can be attributed to random fluctuations.} 
\end{figure}



\begin{figure}
\centering{\includegraphics[width=0.65\textwidth]{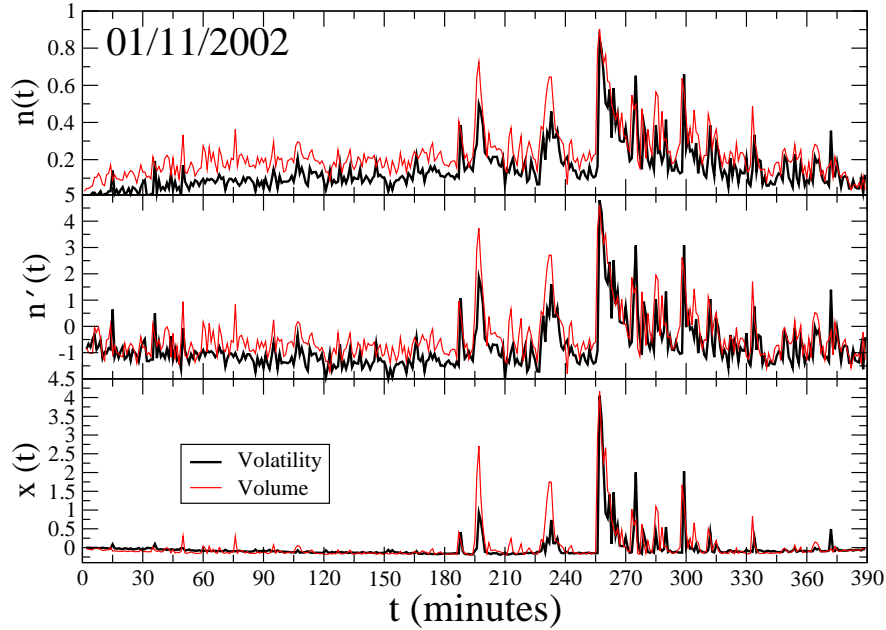}}
  \caption{  (color online)  Example of market co-movement $n(t)$ in both price volatility and total volume, and the qualitative relationship
between the quantities $n(t)$, $n'(t) \equiv (n(t)-\overline{n}(t))/\sigma (t)$, and $x(t) \equiv n(t) n'(t)$. The market shock on  01/11/2002  occurred at $T_{c}=256$ in
response to a public comment by the Fed chairman A. Greenspan concerning economic recovery \cite{NYT}.
  \label{L1b} }
\end{figure}

\begin{figure}
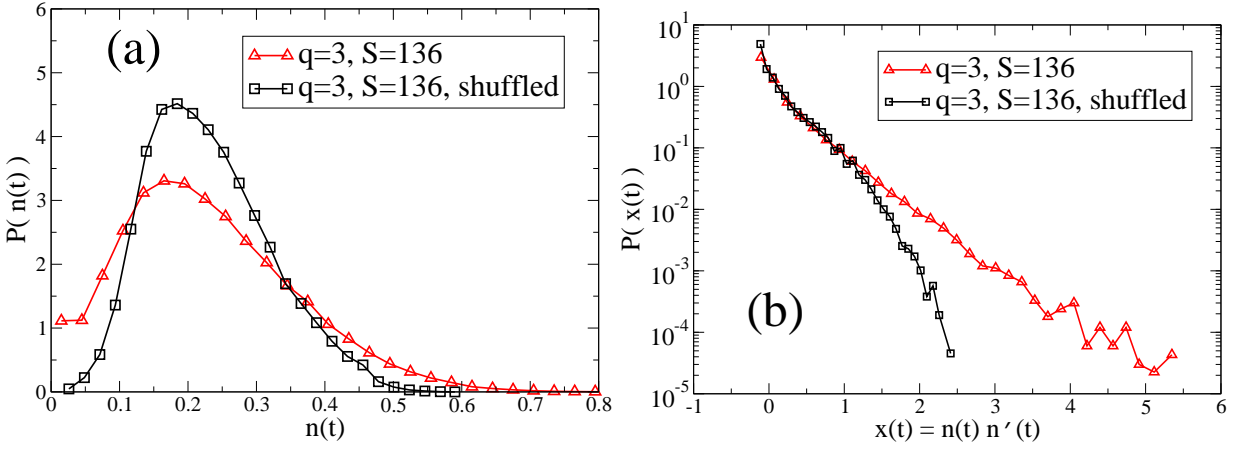

\centering{\includegraphics[width=0.45\textwidth]{Fig5a.eps}}
\centering{\includegraphics[width=0.45\textwidth]{Fig5b.eps}}
 \caption{ (color online)  Using the volatility threshold
$q=3$ and  $S=136$ stocks, we determine the market comovement threshold $x_{c}$ from the pdf of $x(t) \equiv n(t) n'(t)$.
 (a) The  pdf for the 190,000
minutes analyzed of the volatility rate $n(t)$ corresponding to the fraction of the
market with volatility $v_{i}(t) > q$. (b) The pdf of $x(t)$, where in the quantity $x(t)$ we have removed the average daily trend of $n(t)$, so that $x(t)$ is relatively large
when market comovement is large and significant. For comparison, we also plot the pdf of $x_{sh}(t)$ computed from randomly shuffled volatility time series $v_{i}(t)$.
We find a divergence between the pdf of $x(t)$ and of $x_{sh}(t)$ for $x > 1.0$, which we define as the comovement threshold $x_{c} \equiv 1$ in our analysis.
  \label{l1x1pdf} 
}
\end{figure}

\begin{figure}
\centering{\includegraphics[width=0.6\textwidth]{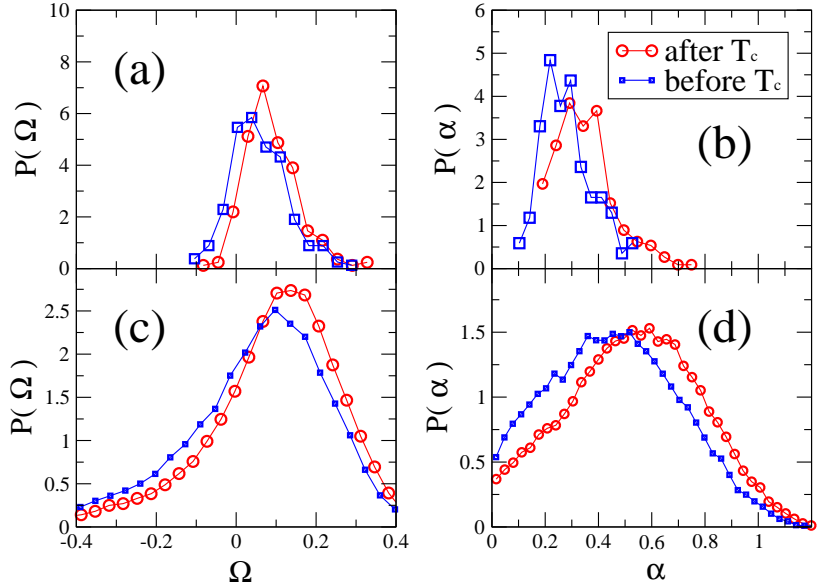}}
\caption{\label{OApdf} (color online) (a,b) Comparison of the probability density functionss $P(\alpha)$ and $P(\Omega)$ of Omori parameters
$\alpha$ and $\Omega$ computed from the average market response $ N_{a,b}(\tau) $.  (c,d) The analogous pdf plots computed from individual stock response
$N^{j}_{a,b}(\tau)$. The average and standard deviation of each data set are (a) $\Omega_{a} = 0.09 \pm 0.07$ ,
$\Omega_{b} = 0.06 \pm 0.07$ (b) $\alpha_{a} = 0.35 \pm 0.11$ , $\alpha_{b} = 0.28 \pm 0.09$ (c) $\Omega_{a} = 0.08 \pm
0.20$ , $\Omega_{b} = 0.03 \pm 0.22$ and (d) $\alpha_{a} = 0.53 \pm 0.25$ , $\alpha_{b} = 0.46 \pm 0.24$.  Values of
both $\Omega_{a}$ and $\alpha_{a}$ are consistently larger than $\Omega_{b}$ and $\alpha_{b}$, indicating that the
response time  after $T_{c}$ is shorter than the activation time leading into $T_{c}$. However the response cascade after $T_{c}$ has,
generally, a larger amplitude.  } 
\end{figure}

\begin{figure}
\centering{\includegraphics[width=0.6\textwidth]{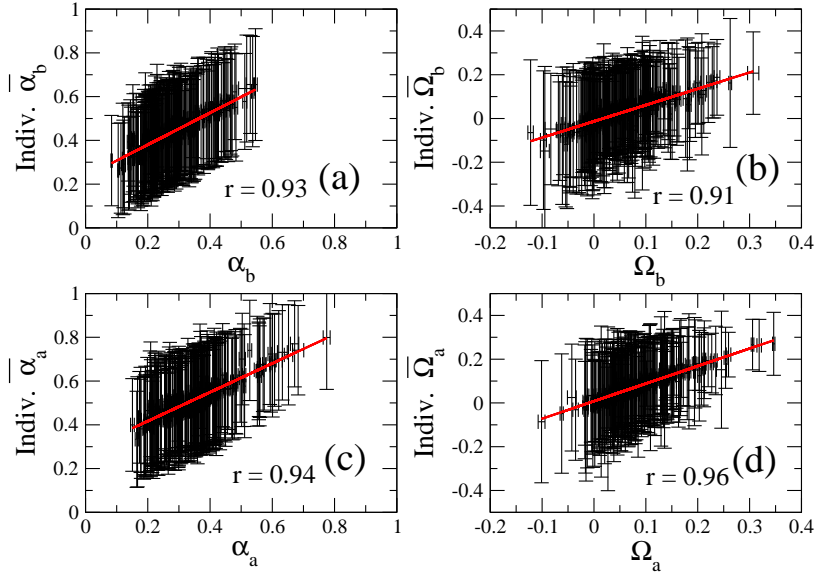}}
\caption{\label{PortInd} (color online)  In order to account for the  dispersion in the pdfs plotted in Figs.~\ref{OApdf}(c) and \ref{OApdf}(d)
for individual stocks, we compare the average values   $\overline{\alpha}_{a,b}$ and $\overline{\Omega}_{a,b}$ computed from all $N^{j}_{a,b}(\tau)$ with
the $\alpha_{a,b}$ and $\Omega_{a,b}$  computed from the corresponding average market response $ N_{a,b}(\tau) $ for each of the 219 $T_{c}$. The visually apparent  correlation  indicates that the parameters quantifying $N_{a,b}(\tau)$ are a good representation of the average $N^{j}_{a,b}(\tau)$.  
The  correlation coefficient $r$ for each  linear regression  is provided in each panel.} 
\end{figure}

\begin{figure}
\centering{\includegraphics[width=0.6\textwidth]{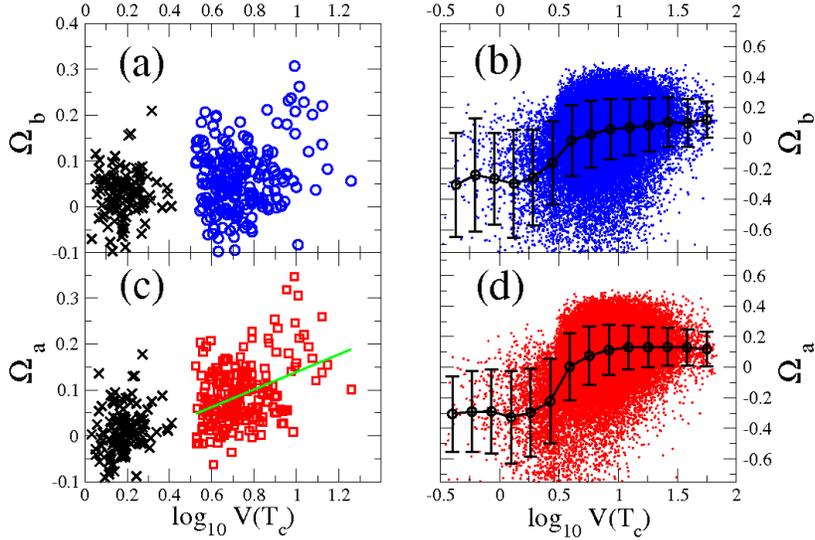}}
\caption{\label{OV} (color online)  The relation between the magnitude $M \equiv \log_{10} V(T_{c})$ and the Omori exponents
$\Omega_{a,b}$. In panels (a) and (c)
 we compare values calculated from the average market response $ N_{a,b}(\tau) $, and in panels (b) and (d) we compare values calculated from  individual stock response
$N^{j}_{a,b}(\tau)$. (a) Weak relation before $T_{c}$, where we validate the linear regression model at $p=0.001$ significance level, but with
correlation coefficient $r=0.22$. The dispersion may result from the variability in anticipation preceding the market
shock at $T_{c}$. (c) The relation between $\Omega_{a}$ and $M$ is stronger after $T_{c}$ than before $T_{c}$, with linear regression significance
$p\approx 0$, correlation $r=0.40$, and regression slope $m= 0.19 \pm 0.03$. The increasing trend demonstrates 
that a faster response, quantified by larger $ \Omega_{a}$, follows a larger  $M$. 
Data points in panels (a) and (c) denoted by the symbol ${\mathsf x }$ correspond to values of $\Omega_{a,b}$ calculated for
randomly selected $T_{c}$ on those 118 days analyzed without a single value of $x(t) > x_{c}$. In panels (b) and (d) 
there is much dispersion in the $\Omega$ values of individual stocks for given $V(T_{c})$. However, the average trends demonstrate a significant crossover at
$M_{\mathsf x}\approx 0.5$ from $\Omega_{a,b} <0$ to $\Omega_{a,b}>0$. The case of $\Omega <0$ can occur when there is more
volatility clustering  for large $\tau$ than for small $\tau$, whereas the case of 
$\Omega>0$ occurs for large volatility cascading around $\tau \gtrsim 0$. This crossover could result from the
difference between anticipated and surprise shocks at $T_{c}$. } 
\end{figure}

\begin{figure}
\centering{\includegraphics[width=0.6\textwidth]{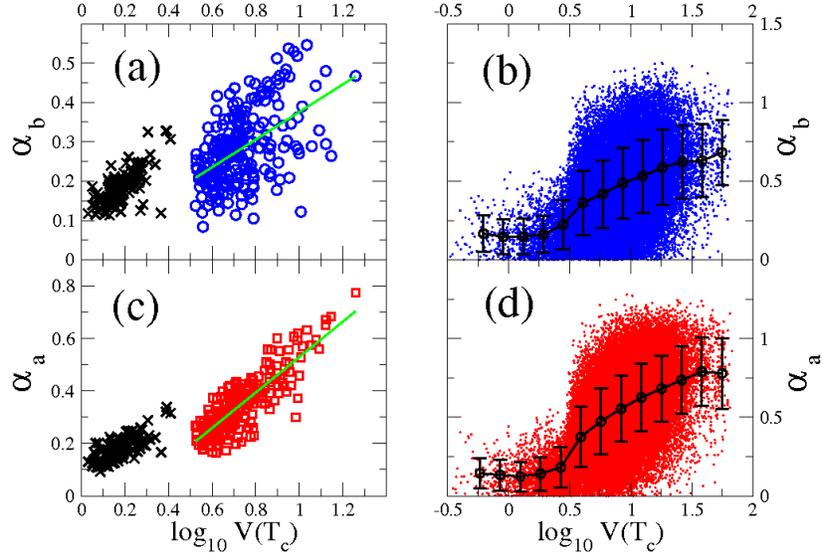}}
\caption{\label{AV} (color online)  The relation between the magnitude $M \equiv \log_{10} V(T_{c})$ and the Omori amplitudes
$\alpha_{a,b}$. In panels (a) and (c) we compare the values calculated from the  average market response $ N(\tau) $ and in panels (b) and (d) we compare values calculated from individual stock response
$N^{j}_{a,b}(\tau)$. (a) The increasing relation between $\alpha_{b}$ and $M$ is statistically stronger than the
relation between $\Omega_{b}$ and $M$ in Fig. \ref{OV}(a), with significance $p\approx 0$, correlation coefficient $r=0.52$ and
regression slope $m= 0.35 \pm 0.04$.  (c) The relation between $\alpha_{a}$ and $M$ is strong, with significance
$p\approx 0$, $r=0.84$, and regression slope $m= 0.68 \pm 0.03$. 
Data points in panels (a) and (c) denoted by the symbol ${\mathsf x }$ correspond to values of $\alpha_{a,b}$ calculated for
randomly selected $T_{c}$ on those 118 days analyzed without a single value of $x(t) > x_{c}$. The result that
$\alpha$ increases with increasing $V(T_{c})$ holds even for random times.
In panels (b) and (d) 
there is much dispersion in the $\alpha$ values of  individual  stocks for given $V(T_{c})$. However the average trends
demonstrate a significant crossover at $M_{\mathsf x}\approx 0.5$ from $\alpha_{a,b} \approx 0.2$ for $M<0.5$ to $\alpha_{a,b}>
0.2$ for $M>0.5$.  This crossover occurs at a similar location as the crossover observed in Figs. \ref{OV}(b) and (d) for
$\Omega_{b,a}$. 
The average amplitude value  $\overline \alpha$ increases sharply for $M > M_{\mathsf x}$, consistent with first order phase
transition behavior. }
\end{figure}

\begin{figure}
\centering{\includegraphics[width=0.6\textwidth]{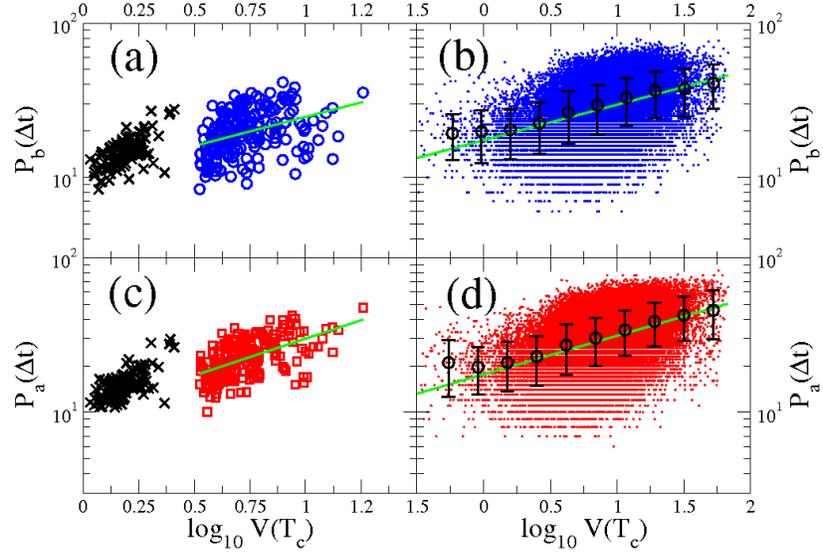}}
\caption{\label{PV} (color online)  The increasing relation between  the productivity $P_{a,b}(\Delta t)$ of each market shock and
the size of the main shock $M \equiv \log_{10} V(T_{c})$ with $\Delta t \equiv 90$ min. As is found in earthquakes, we
find a power-law relationship between $M$ and $V(T_{c})$ described by a productivity exponent $\Pi_{b}$ before
and exponent $\Pi_{a}$ after the market shock. 
Data points in panels (a) and (c) denoted by the symbol ${\mathsf x }$ correspond to values of $P_{a,b}(\Delta t)$ calculated for
randomly selected $T_{c}$ on those 118 days analyzed without a single value of $x(t) > x_{c}$. The result that
$P(\Delta t)$ increases with increasing $V(T_{c})$ holds  even for random times. For the average market response  $
N_{b,a}(\Delta t) $, we find (a) $\Pi_{b} = 0.38 \pm 0.07$ and (c) $\Pi_{a}= 0.48 \pm 0.04$. For the productivity
of individual stocks corresponding to $N^{j}_{b,a}(\Delta t)$ we find (b) $\Pi_{b}= 0.23 \pm 0.01$ and (d) $\Pi_{a} =
0.25 \pm 0.01$. For comparison, the power-law exponent value pertaining to earthquake aftershocks is $\Pi_{a} \approx
0.7-0.9$ \cite{SmallEarthquakes}. }
\end{figure}

\begin{figure}
\centering{\includegraphics[width=0.6\textwidth]{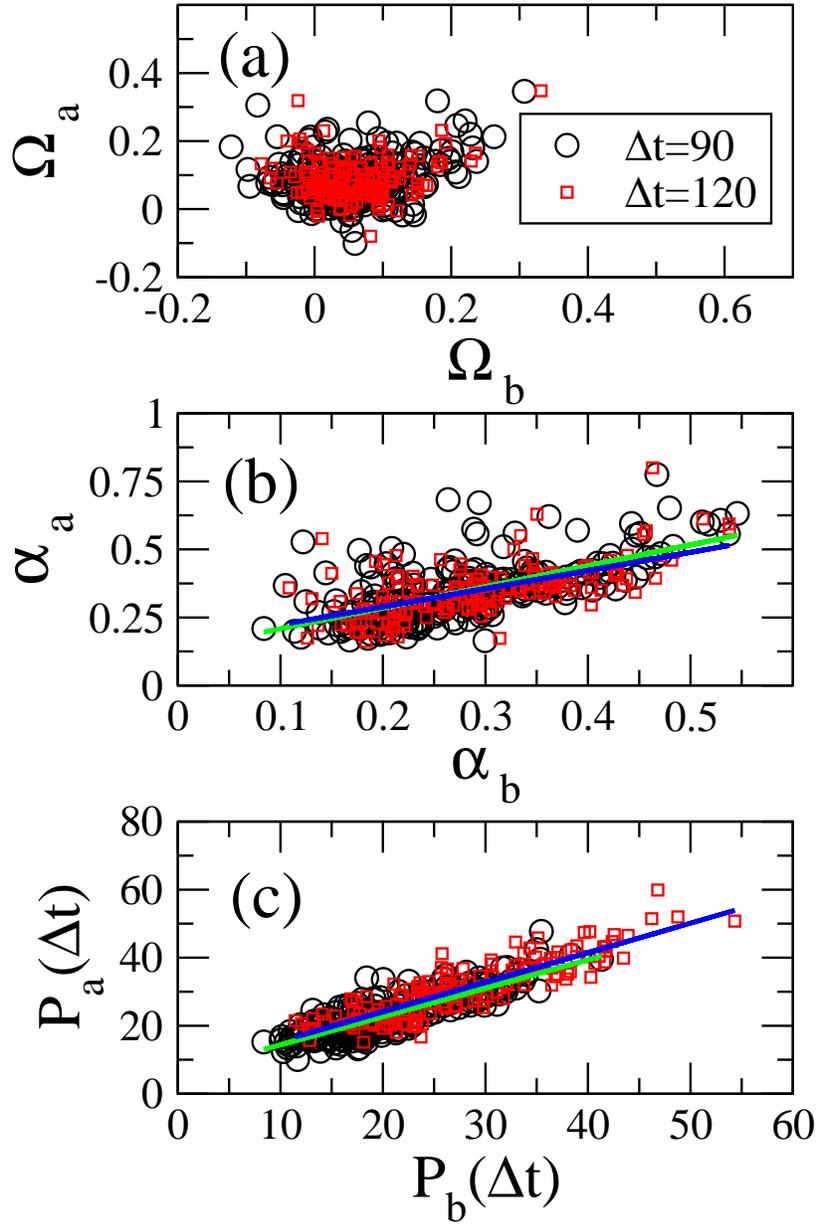}}
\caption{\label{AOPba} (color online)  A comparison of Omori parameters before and after $T_{c}$  for $ N(\tau) $ and
varying $\Delta t$ indicate that $\alpha_{b}$ and $P_{b}(\Delta t)$ are better conditional estimators for the dynamics
after $T_{c}$. (a) Weak relationship between  $\Omega_{b}$ and $\Omega_{a}$ for $\Delta t = 90$ and $120$.
 (b) Strong relationship between  $\alpha_{b}$ and $\alpha_{a}$ for $\Delta t = 90$ and $120$, with both linear
regressions passing the  ANOVA F-test at the $p<0.001$ confidence level.
 (c) Strong relationship between  $P_{b}(\Delta t)$ and $P_{a}(\Delta t)$ for $\Delta t = 90$ and $120$ min. at the $p<0.001$
confidence level.} 
\end{figure}

\begin{figure}
\centering{\includegraphics[width=0.6\textwidth]{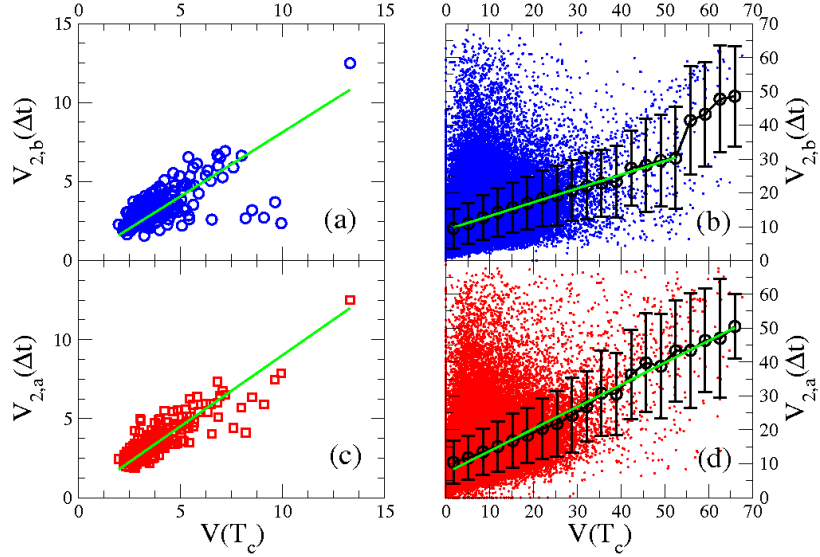}}
\caption{\label{Bath} (color online)  The increasing relation between the size of the main shock $V(T_{c})$ and the size of the second
largest aftershock (or preshock) $V_{2}(\Delta t)$ within $\Delta t$ minutes of $T_{c}$ demonstrates that the volatility of
the largest aftershock (or preshock) increases with mainshock volatility. As with  the Bath law for earthquakes, we observe
a proportional relation $V_{2,a}(\Delta t) \equiv C_{B} V(T_{c})$ which corresponds to a Bath parameter $B=
-\log_{10} C_{B}$. For  the average market response  $ N_{b,a}(\Delta t) $ we calculate $C_{B}$ for (a) the
dynamics before,  $C_{B} = 0. 81$ with correlation coefficient $r=0.70$ and $\chi^{2}= 212$, and for (c) the dynamics
after  $C_{B}= 0.9$ with $r = 0.87$ and $\chi^{2}= 109$. For the Bath law  corresponding to individual stocks  we
find that a linear function best fits the relation between $V(T_{c})$ and the average value $\overline{V}_{2}(\Delta t)$
calculated for equal-sized bins as indicated by circles with one standard deviation error bars. We calculate the
regression slope for the Bath law (b) before is  $m = 0.65 \pm 0.02$ and  (d) after  is $m = 0.40 \pm 0.01$ 
}
\end{figure}

\begin{figure}
\centering{\includegraphics[width=0.6\textwidth]{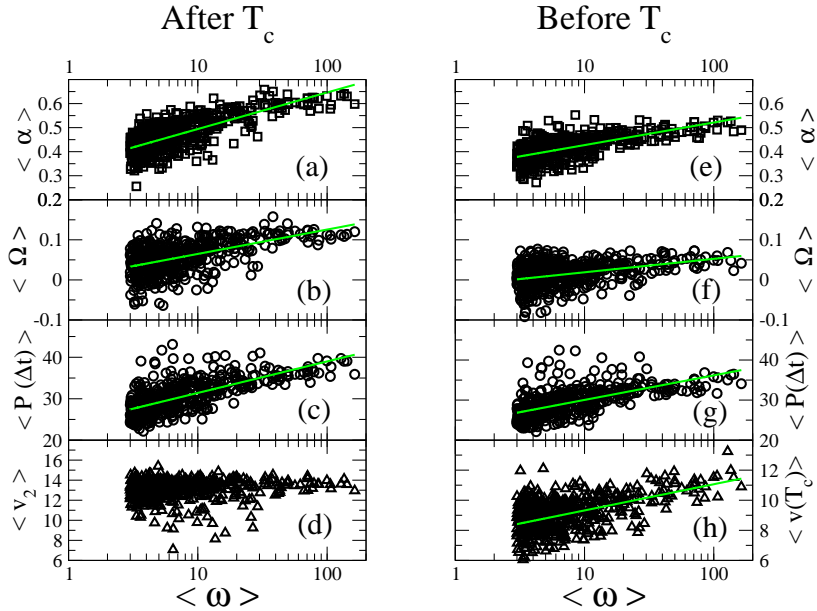}}
\caption{\label{POAV} (color online)  Relations between individual stock trading activity and dynamic response parameters (a-d) after
$T_{c}$ and (e-h) before $T_{c}$, averaged over all the days with a market shock.
 We measure the  trading activity  $ \langle \omega \rangle$  for each stock, defined as the average number of transactions per minute
over the 2-year period 2001-2002. 
 We find that stocks with large trading activity react both more strongly (larger $\alpha$ and larger $P(\Delta t)$) and quickly (larger $\Omega$) to market shocks. 
 However, panel (d) shows that there is little relation between  $ \langle \omega \rangle$  and the average size of the largest aftershock $\langle v_{2} \rangle$.}
\end{figure}
\end{widetext} 
\end{document}